\documentstyle[aps,epsf,psfig]{revtex}
\begin{document}
\draft
\flushbottom

\title{
Charge instabilities near a Van Hove singularity}
\author{J. Gonz\'alez \\}
\address{
        Instituto de Estructura de la Materia. 
        Consejo Superior de Investigaciones Cient{\'\i}ficas. 
        Serrano 123, 28006 Madrid. Spain.}
\date{\today}
\maketitle
\widetext
\begin{abstract}
The charge instabilities of electron systems in the square
lattice are analyzed near the Van Hove singularity by means of
a wilsonian renormalization group approach. We show that the
method preserves the spin rotational invariance at all scales,
allowing a rigorous determination of spin and
charge instabilities of the $t-t'$ Hubbard model. For $t'$ above
$\approx 0.276 t$, repulsive interactions fall into two
different universality classes. One of them has nonsingular
response functions in the charge sector, while the other is
characterized by the splitting of the Van Hove singularity. 
At the level of marginal
perturbations, the Hubbard model turns out to be at the boundary
between the two universality classes, while extended 
models with nearest-neighbor repulsive interactions belong to
the latter class. In the case of open systems allowed to
exchange particles with a reservoir, we show the existence of a
range of fillings forbidden above and below the Van Hove
singularity. This has the property of attracting the Fermi level in
the mentioned range, as the system reaches its lowest energy
when the Fermi energy is at the singularity.

\end{abstract}
\pacs{             }

\tightenlines

\section{Introduction}

The effect of a Van Hove singularity near the Fermi surface of
the CuO layers has been invoked recurrently to understand the
unconventional properties of the high-$T_c$ 
materials\cite{early,rev}. There have been several weak-coupling 
analysis of two-dimensional (2D) models of the
Van Hove singularity, which have shown in particular that the
$t-t'$ Hubbard model may have a phase of $d$-wave
superconductivity\cite{dwave,pin,kohn}. 
The main problem that faces this proposal is
that, although the system is likely to develop strong
antiferromagnetic or superconducting correlations, the effective
interactions grow large at low energies, so that it is not
possible to discern rigorously the ground state of the model. A
related issue concerns the fact that the superconducting
correlations are enhanced like $\log^2
\varepsilon $, when the electron degrees of freedom are 
integrated out down to energy $\varepsilon $ near the Fermi
surface. Recently, some understanding of the system has been
attained by the use of refined renormalization group (RG)
methods\cite{jpn,ren,ren2}.  
The analysis of the low-energy dynamics becomes then
quite subtle, as the Fermi energy has proven to be a dynamical
quantity susceptible itself of renormalization\cite{np,prl}.

The main purpose of this paper is to study the dynamics of the
Fermi surface near a Van Hove singularity. Actually, the
possible relevance of the strong correlations in the system
could be objected by the need of a very fine adjustment of the
Fermi energy at the singularity. We will show, however, that
when the system is allowed to exchange particles with a 
reservoir it finds energetically more favorable to have the 
levels filled up to the position of the singularity. This leads 
to a natural pinning mechanism of the Fermi level over a certain 
range of fillings\cite{mark,pin}.

There is another effect that may be important, at fixed number
of particles. It has been shown by Halboth and Metzner that
the $t-t'$ Hubbard model at the Van Hove filling should have an
instability in its Fermi line leading to a spontaneous breakdown
of the point group symmetry\cite{hm}.  
We will reproduce this effect in
the form of a splitting of the levels of the two inequivalent
saddle points of the 2D band, as a result of the
renormalized interactions between electrons in the two ^^ hot
spots'. In general, we will show that the RG flows in the
charge sector allow to distinguish two different universality
classes for 2D electron systems near a Van Hove
singularity. In one of them, the response functions do not show
any instability under charge perturbations, while in the other
the stable charge distribution is attained after the splitting 
of the Van Hove singularity. We will see, for instance, that
extended Hubbard models with nearest-neighbor repulsive 
interactions and sizeable next-nearest-neighbor hopping belong
to the latter universality class.

\begin{figure}
\begin{center}
\mbox{\epsfysize 7cm \epsfbox{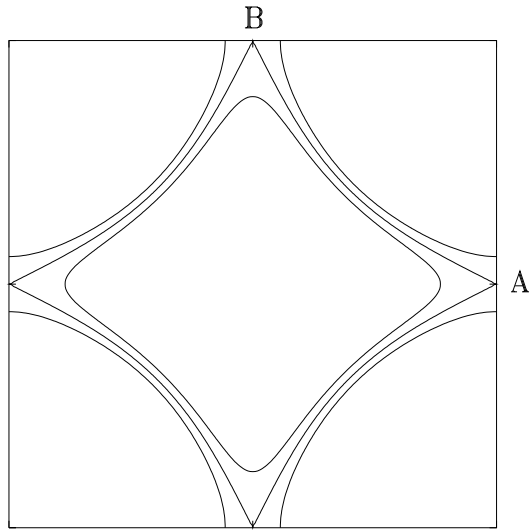}}
\end{center}
\caption{Contour energy map for the $t-t'$ Hubbard model
about the Van Hove filling.}
\label{one}
\end{figure}

Our starting point will be a 2D model of electrons 
in the square lattice 
with nearest-neighbor hopping $t$ and next-nearest-neighbor
hopping $t'$. RG methods are most convenient for the description
of the low-energy behavior of the interactions near the Van Hove
singularity. In the RG approach, high-energy and low-energy
electron modes are separated by an energy cutoff $\Lambda$, that is
sent progressively towards the Fermi line as high-energy modes
are integrated out in the RG process\cite{rg1,rg2}. 
When the Fermi level is at
the Van Hove singularity, as shown in Fig. \ref{one}, most part
of the low-energy states close to the Fermi line are
concentrated around the saddle points at $(\pi, 0)$ and $(0,
\pi)$, as these features are at the origin of the divergent
density of states. Therefore, in building up the low-energy
effective theory we may focus on two patches around the points
$A$ and $B$, where the dispersion relation can be approximated
by
\begin{equation}
\varepsilon_{A,B} ( {\bf k} ) \approx \mp ( t \mp 2 t' ) k_x^2 a^2
\pm ( t \pm 2 t' ) k_y^2 a^2
\end{equation}
$a$ being the lattice constant. From the RG point of view, 
the rest of modes far from the saddle points are irrelevant 
in the continuum limit $a \rightarrow 0$.

In fact, the
effective action for the low-energy modes restricted to the
region $|\varepsilon_{\alpha} \; ({\bf k})| \leq \Lambda $ can be
written in the form
\begin{eqnarray}
S  & = & \int d \omega d^2 k \sum_{\alpha,\sigma} \left( \omega
\; a^{+}_{\alpha,\sigma}({\bf k}, \omega )
   a_{\alpha,\sigma}({\bf k}, \omega ) - \varepsilon_{\alpha} (
{\bf k}) \; a^{+}_{\alpha,\sigma}({\bf k}, \omega )
a_{\alpha,\sigma}({\bf k}, \omega ) \right)   \nonumber   \\ &
&  - U \int d\omega d^2 k \; \rho_{\uparrow } ({\bf k}, \omega)
\; \rho_{\downarrow } (-{\bf k}, -\omega)
\label{actk}  
\end{eqnarray}
where $a_{\alpha,\sigma} (a^{+}_{\alpha,\sigma})$ are electron
annihilation (creation) operators ($\sigma$ labels the spin)
and $ \rho_{\uparrow}, \rho_{\downarrow}$ are the electron density
operators. Under a change in the cutoff $\Lambda
\rightarrow s \Lambda$, with a corresponding scaling of 
the frequency $\omega \rightarrow s \omega$ and
the momenta ${\bf k}
\rightarrow s^{1/2} {\bf k}$, one can check that the effective 
action remains scale invariant after an appropriate scale
transformation of the electron modes, $a_{\alpha,\sigma}
\rightarrow s^{-3/2} a_{\alpha,\sigma}$ \cite{pin}.

In writing the effective action (\ref{actk}) we have taken a
local density-density interaction, like that of the Hubbard
model. A most important point, however, is that in the process
of renormalization other effective interactions may be generated
as well, as long as they are compatible with the symmetries of
the model. This issue will be reviewed in Section 2, ending up
with the proof that our wilsonian RG scheme preserves the spin
rotational invariance.  Section 3 will be devoted to study the
stability of the different distributions of the charge between
the two ^^ hot spots', taking into account the behavior of the
renormalized interactions.  The stability of the location of the
Fermi level around the Van Hove singularity will be discussed in
Section 4, when the system is placed in contact with a charge
reservoir. Finally, Section 5 will be devoted to conclusions and
to comment on possible experimental realizations of our results.

\section{Wilsonian renormalization group}

The wilsonian RG approach, that has been recently applied to the
investigation of many-body electron systems\cite{rg1,rg2}, 
provides a very
efficient way of extracting the effective interactions of the
low-energy theory. It represents an alternative to dealing with
any kind of diagrammatic approximation built from the
effective action (\ref{actk}), which has to suffer from severe
infrared divergences. It is well-known that the different
susceptibilities of the model show logarithmic dependences on
the cutoff $\Lambda$. In the case of the particle-hole
susceptibility $\chi_{ph}({\bf p})$ and the particle-particle
susceptibility $\chi_{pp}({\bf p})$ at small momentum ${\bf p}$,
and the particle-hole susceptibility $\chi_{ph}({\bf q})$ at
${\bf q} \approx {\bf Q} \equiv (\pi , \pi) $, we have\cite{lh}

\begin{eqnarray}
\chi_{ph}  ({\bf p}) & \approx &
\frac{c}{2 \pi^2 t} \log\vert
  \frac{\Lambda}{\varepsilon ({\bf p})}\vert  \label{ph}  \\
\chi_{pp}  ({\bf p})  & \approx &
\frac{c}{4 \pi^2 t} \log^2\vert
  \frac{\Lambda}{\varepsilon ({\bf p})}\vert    \\
\chi_{ph} ({\bf p} + {\bf Q}) & \approx &
\frac{c'}{2 \pi^2 t} \log\vert
  \frac{ \Lambda}{ta^2 {\bf p}^2}\vert
\label{pp}
\end{eqnarray}
where $c \equiv 1/\sqrt{1 - 4(t'/t)^2}$ and $c' \equiv
\log \left[ \left(1 + \sqrt{1 - 4(t'/t)^2} \right)/(2t'/t) \right]$

When performing a RG calculation in the field theory approach,
one computes the variation of the couplings under scale
transformations by taking the
derivatives of the above objects with respect to the cutoff.
The feasibility of the RG method comes from the fact that, in
general, the derivatives of the divergent diagrams do not depend
themselves on the cutoff, what leads to the notion of scaling.
In the present case, however, the derivative of the
particle-particle susceptibility produces a contribution of the
form $\log \vert \Lambda / \varepsilon ({\bf p}) \vert $.  This
leads to an ill-defined computational procedure, as the argument
of the logarithm requires an external {\it ad hoc} parameter for
its definition. Otherwise stated, operators which receive
contributions from particle-particle diagrams display, in
general, cutoff dependences multiplied by a nonlocal, infrared
divergent function of the external momenta\cite{np}. 
This is the
fundamental problem when one tries to apply the RG program to
the model of the Van Hove singularity, which, at present, seems
to find a solution only by promoting the Fermi energy to a
renormalized, scale-dependent variable\cite{np,prl}.

Opposite to the field theory RG approach, the wilsonian RG
approach provides a better computational framework to deal with
the above problem, as it makes a clear distinction of the
operators which are renormalized in the particle-particle
channel. The idea is to find the low-energy effective theory by
identifying the operators that scale appropriately as the cutoff
is sent to zero.  This task is accomplished by performing a
progressive integration of high-energy modes living in two thin
shells of width $d \Lambda $, at distance $\Lambda $ in energy
below and above the Fermi surface.  In this process one keeps
only operators which remain scale invariant, or which receive
corrections at most of order $d \Lambda $, as the rest
of contributions vanish in the limit 
$\Lambda \rightarrow 0$ \cite{rg1}.

\begin{figure}
\begin{center}
\mbox{\epsfysize 7cm \epsfbox{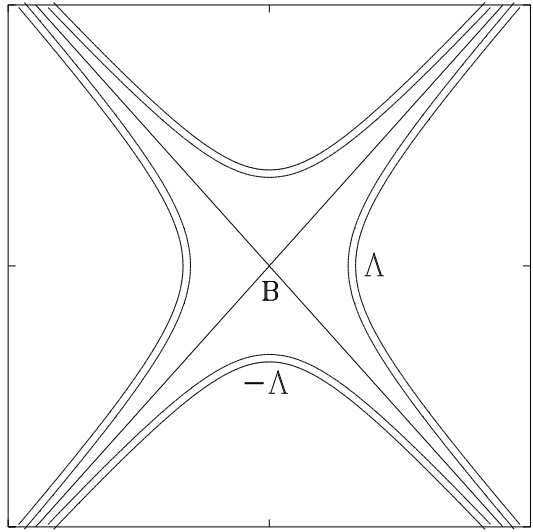}}
\end{center}
\caption{Plot of the slices with high-energy states integrated
in the renormalization group process.}
\label{two}
\end{figure}

\begin{figure}
\begin{center}
\mbox{\epsfysize 3cm \epsfbox{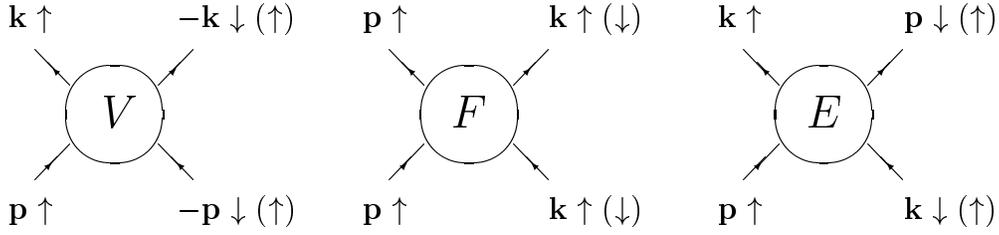}}
\end{center}
\caption{Different channels that undergo renormalization in
the wilsonian approach.}
\label{three}
\end{figure}

Let us concentrate on the region around one of the saddle
points, in which the two thin slices of width $d \Lambda $ look
as shown in Fig. \ref{two}. The modes in the two slices build up
the intermediate states in the corrections by particle-hole
and particle-particle diagrams to the vertex functions of the
theory. Focusing on the four-point function, we observe that
such corrections are linear in  $d \Lambda $ only in a reduced
number of instances. Actually, contributions of order $d \Lambda
/ \Lambda $ arise for the same kinematics which do not make
irrelevant the four-point function in Fermi liquid theory\cite{rg1}. 
They amount to three different possibilities, which are represented
graphically in Fig. \ref{three}.

\begin{figure}
\begin{center}
\mbox{\epsfxsize 3cm \epsfbox{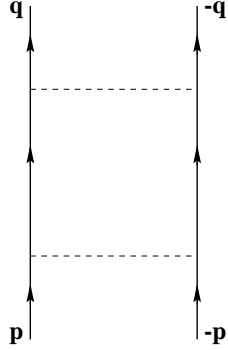}}
\end{center}
\caption{Particle-particle diagram renormalizing the BCS channel
at the one-loop level.}
\label{four}
\end{figure}

The BCS channel, that we denote by $V$, opens up when the
momenta of the incoming particles add up to zero. At the
one-loop level, for instance, it receives a contribution from
the particle-particle diagram in Fig. \ref{four}. It is clear
that, for each internal line with momentum ${\bf k}$ in the
slice of width $d \Lambda $, the opposite momentum ${- \bf k}$
is also found among the high-energy states integrated over, so
that the diagram is of order $\sim d \Lambda $. A similar
argument shows that, when the sum of the momenta of the incoming
particles is not zero, the set of available intermediate states
is reduced to the intersection of two slices, displaced with
respect to each other, and the phase space to build the diagram
becomes of order $\sim (d \Lambda)^2 $.  Thus, in the wilsonian
RG approach the particle-particle diagram only renormalizes the
vertex function for the precise kinematics of the BCS channel,
while it produces irrelevant contributions for other choices of
the external momenta\cite{rg1}.

The forward-scattering channel $F$ is singled out in the vertex
function when the momentum transfer along one of the fermion
lines vanishes in the diagram. Technically, we may distinguish
it from the exchange channel $E$, which arises when the momentum
transfer between two lines connected by the interaction
vanishes. It is clear, though, that when the incoming and
outgoing particles have all the same spin the respective
channels $F_{\parallel}$ and $E_{\parallel}$ contribute with
opposite sign to the same scattering amplitudes. It can be
checked that all the corrections can be written in terms of the
combination $F_{\parallel} - E_{\parallel}$, so that
$E_{\parallel}$ can be redefined away by introducing the
coupling $\tilde{F}_{\parallel} = F_{\parallel} -
E_{\parallel}$.

\begin{figure}
\begin{center}
\mbox{\epsfysize 4cm \epsfbox{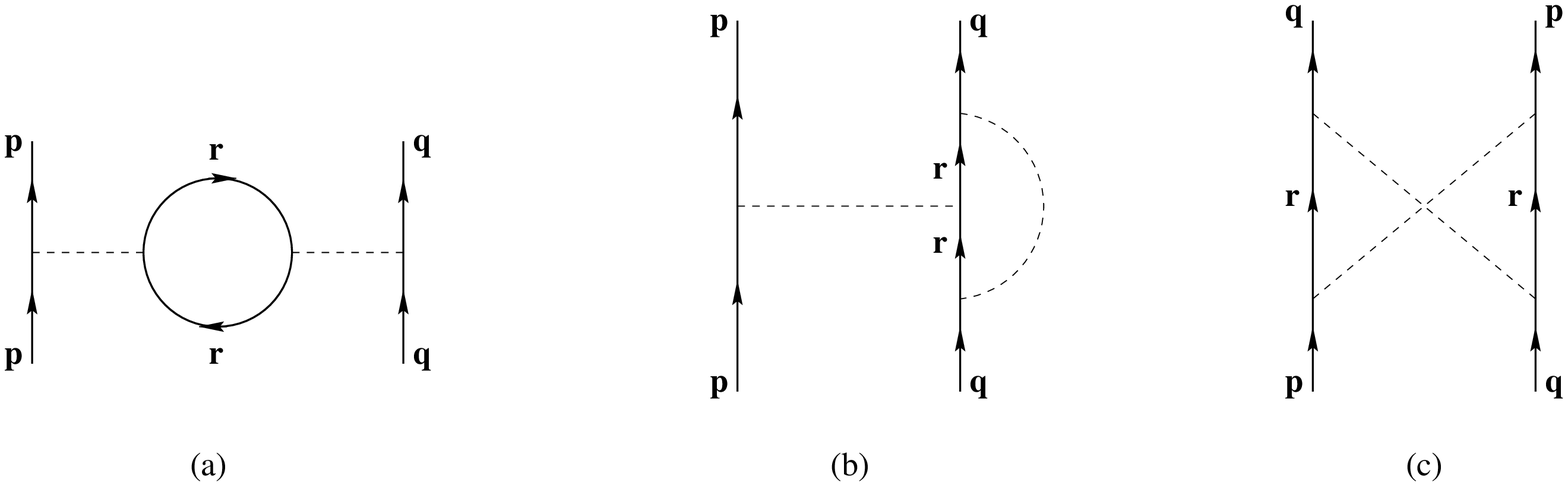}}
\end{center}
\caption{Particle-hole diagrams renormalizing the $F$ channel
at the one-loop level.}
\label{five}
\end{figure}

For incoming and outgoing particles with the same spin, 
all the diagrams shown in Fig. \ref{five}, with internal
momenta in the slices of width $d \Lambda $ integrated over,
produce a correction of order $\sim d \Lambda $ to the 
$\tilde{F}_{\parallel}$ coupling. Similarly, diagrams (a)
and (b) in Fig. \ref{five} are responsible for a renormalization
of order $\sim d \Lambda $ of the $F_{\perp}$ coupling.
This is a
consequence of the fact that, for no matter how small momentum
transfer, one can always build particle-hole excitations in the
asymptotic region where two slices approach the Fermi line, as
observed in Fig. \ref{two}.

\begin{figure}
\begin{center}
\mbox{\epsfxsize 3cm \epsfbox{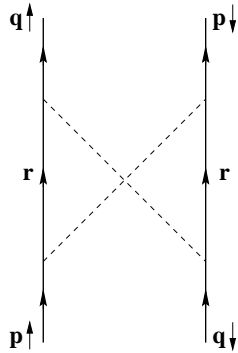}}
\end{center}
\caption{Particle-hole diagram renormalizing the $E_{\perp}$
channel at the one-loop level.}
\label{six}
\end{figure}

On the other hand, when the
vanishing momentum transfer takes place from one particle to
another with different spin, we may still think of it as a
different channel, that we call the $E_{\perp}$ exchange
channel\cite{ex}. 
In that case, a number of intermediate particle-hole
excitations of order $\sim d \Lambda $ can be counted from the
diagram in Fig. \ref{six}, which is the only one that
renormalizes the $E_{\perp}$ channel.

It can be appreciated from the diagrams in Figs. \ref{five} and
\ref{six} that the renormalization of the $F$ channel only depends 
on the $\tilde{F}_{\parallel}$ and $F_{\perp}$ couplings, as
well as only $E_{\perp}$ couplings enter in the diagrams
renormalizing
the $E_{\perp}$ channel. On the other hand, the analysis of
the instabilities of the model can be carried out in parallel,
by using either set of couplings, together with the $V$
couplings. This is due to the fact that the $F$ couplings feed
the correlations of the $z$-projection of the spin operators,
while the $E_{\perp}$ couplings drive the correlations for the
$x$ and $y$ components. The flow equations for the $F$ and the
$E_{\perp}$ couplings have been studied in Refs. \onlinecite{prl}
and \onlinecite{jpn}, respectively. It can be shown that the
different phases that one obtains for the model (ferromagnetism,
antiferromagnetism, superconductivity) do not depend on the use
of one set of equations or the other.  This relies on the key
assumption of spin rotational invariance of our RG scheme, that
we turn to check next.

It is possible to show that the response functions that measure
the spin correlations for the $x$, $y$ and $z$ components of the
spin are numerically equal, at each point of the RG flow, with a
suitable choice of the bare couplings of the model. We deal in
particular with the response functions at zero momentum, which
measure the correlations of the operators
\begin{equation}
 S_i = \sum_{k} \sum_{\alpha = A,B}
a^{+}_{\alpha,\sigma}({\bf k}) \sigma_{\sigma \sigma '}^i
a_{\alpha,\sigma '}({\bf k}) \;\;\;\;\;\;\; i = x,y,z
\end{equation}
The following analysis can be also applied with complete
similarity to the response functions at finite 
wavevector ${\bf Q} \equiv (\pi , \pi)$.

\begin{figure}
\begin{center}
\mbox{\epsfxsize 4cm \epsfbox{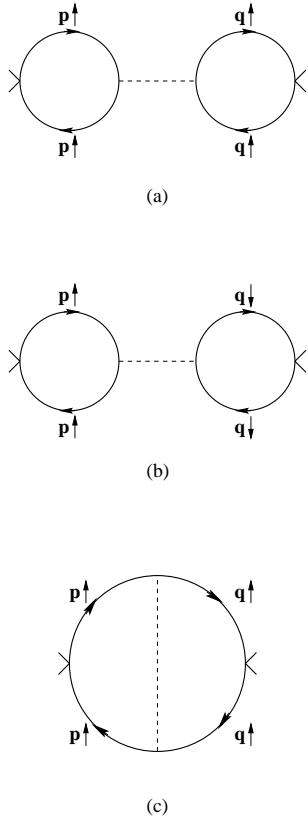}}
\end{center}
\caption{First-order contributions to the correlator of the
$S_z$ operator.}
\label{seven}   
\end{figure}

Scaling equations for the response functions can be derived in
the same fashion as for the renormalizable one-dimensional
models\cite{sch}.  
The first-order contributions to the response function
$R_z (\omega)$ for the $S_z $ operator are given in Fig.
\ref{seven}. We introduce here a distinction between the
interactions $F_{intra}$ and $E_{intra}$ for currents in the
same saddle point and the interactions $F_{inter}$ and
$E_{inter}$ between currents at different saddle points. After
taking the derivative with respect to the cutoff and imposing
the self-consistency of the diagrammatic expansion, we obtain
\begin{equation}
\frac{\partial R_{z}}{\partial \Lambda} = -  \frac{2c}{\pi^2 t} 
  \frac{1}{\Lambda}  +  \frac{c}{ \pi^2 t} \left( \tilde{F}_{intra
\parallel} - F_{intra \perp} + \tilde{F}_{inter \parallel} -
F_{inter \perp}  \right) \frac{1}{\Lambda}  R_{z}
\label{z}
\end{equation} 
where we have used the redefinition $\tilde{F}_{\parallel} \equiv
F_{\parallel} - E_{\parallel}$.

\begin{figure}
\begin{center}
\mbox{\epsfxsize 3cm \epsfbox{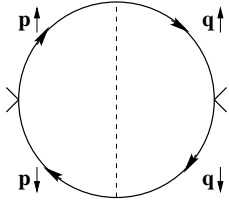}}
\end{center}
\caption{First-order contribution to the correlators
of the $S_x$ and $S_y$ operators.}
\label{eight}
\end{figure}

Similar scaling equations can be obtained for the response
functions for the other projections of the spin, $R_x$ and
$R_y$. In both cases, we have the first-order contribution shown
in Fig. \ref{eight}.  The scaling equation for $R_x$, for
instance, reads
\begin{equation}
\frac{\partial R_{x}}{\partial \Lambda} = -  \frac{2c}{\pi^2 t}
  \frac{1}{\Lambda}  -  \frac{c}{ \pi^2 t} \left( E_{intra \perp}  +
E_{inter \perp} \right) \frac{1}{\Lambda}  R_{x}
\label{x}
\end{equation}

From inspection of Eqs. (\ref{z}) and (\ref{x}), it turns out
that $R_x$, $R_y$ and $R_z$ are identical as long as the
constraints $F_{intra \perp}  - \tilde{F}_{intra \parallel} =
E_{intra \perp}$ and $F_{inter \perp} - \tilde{F}_{inter
\parallel} = E_{inter \perp}$ are fulfilled at all the points of
the RG flow.  It can be easily seen that this is in fact the
case provided that the initial values of the couplings satisfy
both conditions.

The RG flow equations for the interactions in the forward
scattering channel can be obtained from Ref. \onlinecite{prl}.
For the combinations $F_{intra \perp}  - \tilde{F}_{intra
\parallel}$ and $F_{inter \perp} - \tilde{F}_{inter \parallel}$,
they read
\begin{eqnarray}
\Lambda   \frac{\partial \left( F_{  intra \perp} - \tilde{F}_{
intra \parallel} \right)}{\partial \Lambda}  & = & - \frac{1}{2\pi^2
t} c  \left[ \left( F_{ intra \perp} - \tilde{F}_{ intra
\parallel} \right)^2 +  \left( F_{  inter \perp} -
\tilde{F}_{  inter \parallel} \right)^2 \right] \label{flow11} \\
\Lambda   \frac{\partial \left( F_{ inter \perp} - \tilde{F}_{
inter \parallel} \right)}{\partial \Lambda}  & = & -
\frac{1}{\pi^2 t} c
 \left( F_{ intra \perp} - \tilde{F}_{ intra \parallel } \right)
\left( F_{ inter \perp} - \tilde{F}_{ inter \parallel} \right)
\label{flow12}
\end{eqnarray}
The RG equations in the exchange channel can be taken from Ref.
\onlinecite{jpn}. For $E_{intra \perp}$ and $E_{inter \perp}$
they have the form\cite{note}
\begin{eqnarray}
 \Lambda \frac{\partial E_{intra \perp}}{\partial \Lambda}  & = &
 - \frac{1}{2\pi^2 t} c \left( E_{intra \perp}^2 + E_{inter \perp
}^{2}  \right)    \label{flow21}       \\ 
 \Lambda \frac{\partial E_{inter \perp}}{\partial \Lambda }  & = & 
  - \frac{1}{\pi^2 t} c \left(
E_{intra \perp} E_{inter \perp} \right)
\label{flow22}
\end{eqnarray}

It becomes manifest that, if $F_{intra \perp}  -
\tilde{F}_{intra \parallel} = E_{intra \perp}$ and
$F_{inter \perp} - \tilde{F}_{inter \parallel} = E_{inter
\perp}$ at the upper value of the cutoff, the two constraints
are satisfied at any lower scale. Let us remark that this choice
of initial conditions is actually quite reasonable, as it is
what one would make by taking the bare values of the Hubbard
interaction.  We conclude that the spin rotational invariance of
the model is preserved within our RG scheme, what is a rather
remarkable result given the nontrivial flow of the RG equations.
By taking the initial conditions $F_{intra \perp}  -
\tilde{F}_{intra \parallel} > 0$ and
$F_{inter \perp} - \tilde{F}_{inter \parallel} >0$, we observe
that these combinations flow to strong coupling at low energies.
In general, this kind of behavior leads to instabilities in the
spin sector of the model, which have been studied by several
authors\cite{dwave,jpn,ren,ren2}.  

\begin{figure}
\begin{center}
\mbox{\epsfxsize 9cm \epsfbox{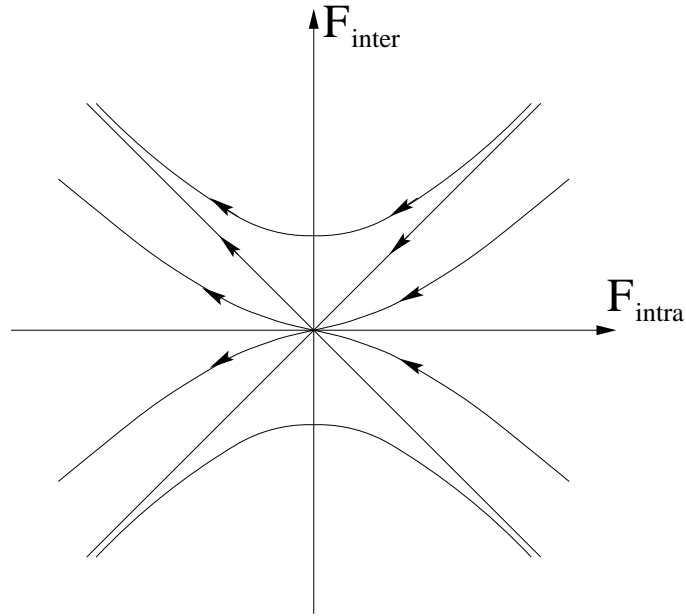}}
\end{center}
\caption{Flow of the renormalized interactions in the
$\left( \tilde{F}_{ intra \parallel} + F_{  intra \perp} ,
  \tilde{F}_{ inter \parallel} + F_{ inter \perp} \right)$
plane.}
\label{ten}
\end{figure}

We have moreover the complementary flow equations
\begin{eqnarray}
\Lambda   \frac{\partial \left( F_{  intra \perp} + \tilde{F}_{
intra \parallel} \right)}{\partial \Lambda}  & = & \frac{1}{2\pi^2
t} c  \left[ \left( F_{ intra \perp} + \tilde{F}_{ intra
\parallel} \right)^2 +  \left( F_{  inter \perp} +
\tilde{F}_{  inter \parallel} \right)^2 \right] \label{flow31}  \\
\Lambda   \frac{\partial \left( F_{ inter \perp} + \tilde{F}_{
inter \parallel} \right)}{\partial \Lambda}  & = &   
\frac{1}{\pi^2 t} c
 \left( F_{ intra \perp} + \tilde{F}_{ intra \parallel } \right)
\left( F_{ inter \perp} + \tilde{F}_{ inter \parallel} \right)
\label{flow32}
\end{eqnarray}
We assume that the bare couplings are such that $F_{  intra
\perp} + \tilde{F}_{ intra \parallel} > 0$ and $F_{ inter \perp}
+ \tilde{F}_{  inter \parallel} > 0$. Under these conditions,
the flow may be attracted to two different regions, which
characterize respective universality classes. When the initial
couplings satisfy $F_{  intra \perp} + \tilde{F}_{ intra
\parallel} > F_{ inter \perp} + \tilde{F}_{ inter \parallel}$,
both combinations are renormalized to zero at low energies. The
complete set of RG equations has then the asymptotic solution
$\tilde{F}_{intra \parallel} \approx - F_{intra \perp}$ and
$\tilde{F}_{inter \parallel} \approx - F_{inter \perp}$.
Otherwise, when $F_{  intra \perp} + \tilde{F}_{ intra
\parallel} < F_{ inter \perp} + \tilde{F}_{  inter \parallel}$
at the initial stage, the flow for these combinations of
couplings becomes unstable, as shown in Fig. \ref{ten}.
The corresponding universality class is characterized by
the asymptotic behavior $F_{  intra \perp} +
\tilde{F}_{ intra \parallel} \approx - (F_{ inter \perp} +
\tilde{F}_{  inter \parallel})$. This leads to important
consequences in the charge sector, as we will see in what
follows.

\section{Charge dynamics between ^^ hot spots'}

\begin{figure}
\begin{center}
\mbox{\epsfxsize 4cm \epsfbox{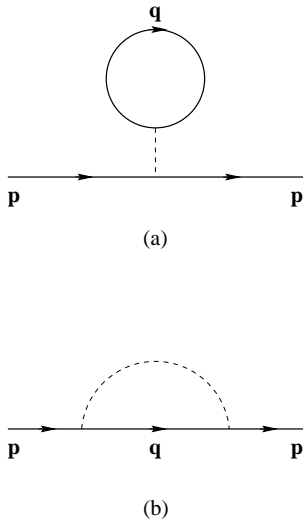}}
\end{center}
\caption{Diagrams contributing to the electron self-energy at
the one-loop level.}
\label{nine}
\end{figure}

The couplings $F_{  intra \perp} + \tilde{F}_{ intra \parallel}$
and $F_{ inter \perp} + \tilde{F}_{ inter \parallel}$ drive the
interactions in the charge sector. They control the way in which
the chemical potential is renormalized in the model. The
chemical potential $\mu$ is introduced to fix the Fermi energy,
but it gets corrections due to the charge present in the
system. At the one-loop level, these corrections come from the
diagrams in Fig. \ref{nine}. The inspection of the kinematics of
these diagrams shows that the charge in the system interacts
through the combination of the couplings $F_{  intra \perp} +
F_{ intra \parallel} - E_{ intra \parallel}$ and $F_{  inter
\perp} + F_{ inter \parallel} - E_{ inter \parallel}$. These are
actually what we have called $F_{  intra \perp} + \tilde{F}_{
intra \parallel}$ and $F_{  inter \perp} + \tilde{F}_{ inter
\parallel}$, respectively.  The sum of all these couplings
renormalizes to zero, in either of the two universality classes
mentioned at the end of Section 2.  This means that, when the
system is considered in isolation, its compressibility cannot be
very different from that of the noninteracting model.

When the model falls in the universality class with the unstable
flow $F_{  intra \perp} + \tilde{F}_{ intra \parallel} \approx -
(F_{ inter \perp} + \tilde{F}_{  inter \parallel})$, a mismatch
in the filling levels of the two ^^ hot spots' $A$ and $B$ may
arise. This has been anticipated by Halboth and Metzner in a
RG study of the $t-t'$ Hubbard model, in the form of a Pomeranchuk
instability of the Fermi line\cite{hm}.  
The same kind of effect can be obtained in our model as an 
instability in the response function $R_{AB}$ to perturbations in 
the difference of charge densities $n_A$ and $n_B$
at the two patches $A$ and $B$. 

A scaling equation for the dynamic correlator $R_{AB} (\omega) $ 
of the operator $n_A - n_B$ can
be derived with the same technique applied in Section 2 to the
spin response functions. We obtain an expression of the
form
\begin{equation}
\frac{\partial R_{AB}}{\partial \Lambda} = -  \frac{2c}{\pi^2 t} 
  \frac{1}{\Lambda}  +  \frac{c}{ \pi^2 t} \left( F_{intra \perp}  +
\tilde{F}_{intra \parallel} - F_{inter \perp} - \tilde{F}_{inter
\parallel} \right) \frac{1}{\Lambda}  R_{AB}
\label{AB}
\end{equation}
From this equation, it can be checked that $R_{AB}$ develops a
divergence at a finite value of the frequency whenever the bare
couplings satisfy $F_{intra \perp}  + \tilde{F}_{intra
\parallel} - F_{inter \perp} - \tilde{F}_{inter \parallel} < 0$.
This is the signal that, when the Fermi level is nominally at
the Van Hove singularity, an excess of charge develops in one of
the ^^ hot spots' over the other.

The precise nature of this instability can be clarified by
performing a self-consistent solution of the Schwinger-Dyson
equation 
\begin{equation}
G^{-1} = G_0^{-1} - \Sigma
\end{equation}
in our model with the two ^^ hot spots'. The Fermi energy 
$\varepsilon_F $ in the full electron Green function $G$ is
determined from the balance between the chemical potential $\mu$ 
in the free electron Green function $G_0$ and the corrections to it
introduced by the electron self-energy. These corrections come
at the one-loop level from the diagrams in Fig. \ref{nine},
which depend in turn on the charge present in the system.
Self-consistency is attained when the chemical potential after
such renormalization matches the highest occupied level.

To study the interaction between the charge in the two ^^ hot
spots', we model each of them by a singular density of states of
the form
\begin{equation}
n (\varepsilon) = - \frac{1}{\Lambda } \log (|\varepsilon |
/\Lambda ) \;  ,  -\Lambda < \varepsilon < \Lambda
\label{dos}
\end{equation}
Furthermore, for the same nominal chemical potential $\mu $ of
the system, we introduce two independent Fermi levels
$\varepsilon_A $ and $\varepsilon_B $ for the respective ^^ hot
spots'. The Schwinger-Dyson equation referred to these two
variables splits then in two equations of the form
\begin{eqnarray}
\varepsilon_{A}  & = &   \mu -  \int^{\varepsilon_{A}}_{-\Lambda} 
d \varepsilon \; ( F_{intra \perp} (\varepsilon ) +
\tilde{F}_{intra \parallel} (\varepsilon ) ) \; n (\varepsilon )  -
\int^{\varepsilon_{B}}_{-\Lambda} d \varepsilon  \; ( F_{inter
\perp} (\varepsilon ) + \tilde{F}_{inter \parallel} (\varepsilon
) )  \;  n (\varepsilon )   \label{self11}     \\
\varepsilon_{B}  & = &   \mu -  \int^{\varepsilon_{B}}_{-\Lambda}
d \varepsilon  \; ( F_{intra \perp} (\varepsilon ) +
\tilde{F}_{intra \parallel} (\varepsilon ) ) \; n (\varepsilon )  -
\int^{\varepsilon_{A}}_{-\Lambda} d \varepsilon  \; ( F_{inter
\perp} (\varepsilon ) + \tilde{F}_{inter \parallel} (\varepsilon
) )  \;  n (\varepsilon )
\label{self12}
\end{eqnarray}
where we have introduced renormalized vertices in place of the
four-point interactions in Fig. \ref{nine}.  We remark that
$\varepsilon_{A}$ and $\varepsilon_{B}$ are measured in the
reference frames in which the dependence of the density of
states is fixed by Eq.  (\ref{dos}). Thus, the fact that
$\varepsilon_{A}$ and $\varepsilon_{B}$ may be nominally
different after renormalization is just a consequence of that
convention. The physical picture is however the opposite, namely
that the one-particle levels are shifted to higher energy by a
different amount in each of the two ^^ hot spots', up to a point
in which the respective Fermi levels reach the common chemical
potential.

\begin{figure}
\begin{center}
\mbox{\epsfxsize 7cm \epsfbox{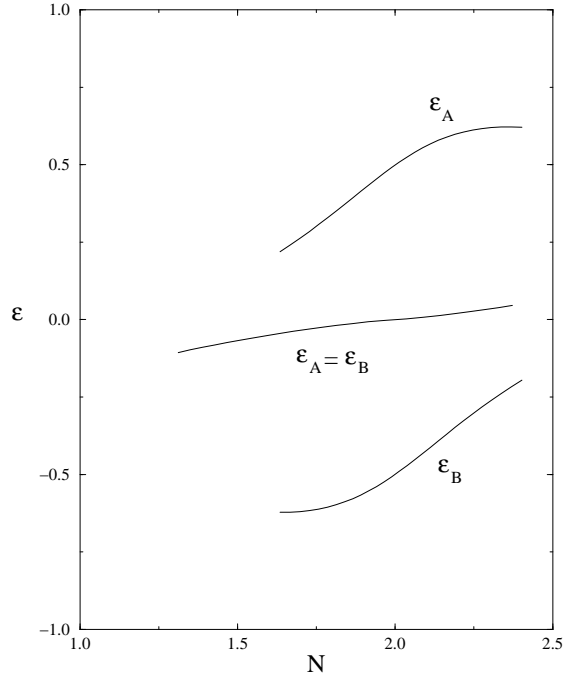}}
\end{center}
\caption{Filling levels in the two ^^ hot spots' versus total
charge $N$.
Middle curve: solution with even distribution of the charge.
Upper and lower curves: uneven filling levels of the solution
corresponding to the splitting of the Van Hove singularity.} 
\label{eleven}
\end{figure}

It can be checked that, in the phase with the stable flow
$F_{intra \perp}  + \tilde{F}_{intra \parallel} > F_{inter
\perp} + \tilde{F}_{inter \parallel}$, the Eqs.
(\ref{self11}) and (\ref{self12}) only admit a single solution
with $\varepsilon_{A} = \varepsilon_{B}$. However, for couplings
falling in the universality class with the unstable flow,
together with that solution we find another which has different
filling levels for the two ^^ hot spots'.  A plot of the filling
levels versus the total charge in the system is represented in
Fig. \ref{eleven}, for the particular bare values $F_{intra
\perp} =  \Lambda$, $F_{inter \perp} = 2 \Lambda$. We have found
that the solution with $\varepsilon_{A} \neq \varepsilon_{B}$
turns out to have always the lowest energy. The physical
interpretation of these results is that, due to the 
mismatch in the repulsive
interaction, the one-particle levels are shifted upwards with
higher strength in one of the ^^ hot spots' than in the other,
so that the common Fermi energy becomes placed below one of the
saddle points and above the other. The lowest-energy solution
describes therefore the splitting of the Van Hove singularity, in
correspondence with the spontaneous breakdown of the
tetragonal symmetry found by Halboth and Metzner.

If one were to take the nominal couplings of the Hubbard model 
as the bare interactions in the RG approach, 
this would lead to the initial condition $F_{intra \perp}
= F_{inter \perp} = U$, with the rest of $F$ couplings equal to
zero. Thus, the Hubbard model is placed right at the boundary
between the region of unstable flow and the phase in which the
$F_{intra \perp}  + \tilde{F}_{intra \parallel}$ and $F_{inter
\perp}  + \tilde{F}_{inter \parallel}$ couplings are
renormalized to zero. The slightest perturbation by any
irrelevant operator may drive the system to either of the two
phases, and the work of Halboth and Metzner\cite{hm} 
shows indeed that
the Hubbard model in particular falls in the universality class
with the charge instability. 

However, the splitting of the Van Hove singularity is not a 
universal feature of 2D electron systems. There may exist models
that lead
instead to the initial condition $F_{intra \perp}  +
\tilde{F}_{intra \parallel} > F_{inter \perp} + \tilde{F}_{inter
\parallel}$. According to Eqs. (\ref{flow31}),
(\ref{flow32}) and (\ref{AB}), these models only have a weak,
nonsingular response to any charge perturbation, and they do not
show therefore the splitting of the Van Hove singularity.

We have to bear in mind that the energies at which the response
to a charge perturbation becomes unstable may be rather small.
From Eq. (\ref{AB}), the instability arises at the point in
which the renormalized coupling $F_{inter \perp} +
\tilde{F}_{inter \parallel} - F_{intra \perp}  - 
 \tilde{F}_{intra \parallel}$ diverges. This happens at a
frequency
\begin{equation}
\omega \approx \Lambda \exp \left\{ - (2\pi^2 t)/
 \left( c ( F_{inter \perp}(\Lambda) + \tilde{F}_{inter
\parallel}(\Lambda) - F_{intra \perp}(\Lambda)  - \tilde{F}_{intra
\parallel}(\Lambda) ) \right)  \right\}
\end{equation}

For the $t-t'$ Hubbard model,
the response functions at vanishing momentum
dominate over those for perturbations with finite wavevector
${\bf Q} \equiv (\pi ,\pi )$ when $c > c'$ \cite{jpn}. 
Taking the expressions of $c$ and $c'$ given after Eqs.
(\ref{ph})-(\ref{pp}), this corresponds to values of $t'$ above 
$\approx  0.276 t$ \cite{note3}.  In this region of
the phase diagram, a strong instability leading to
ferromagnetism should also be present, as it has been shown in
Ref. \onlinecite{sor}. The mismatch between the densities of
spin up and spin down electrons is driven by the coupling
$F_{intra \perp} - \tilde{F}_{intra \parallel} + F_{inter \perp}
- \tilde{F}_{inter \parallel}$. Given that the coupling for the
charge instability, $F_{inter \perp} +
\tilde{F}_{inter \parallel} - F_{intra \perp} - 
\tilde{F}_{intra \parallel}$, 
is nonvanishing in the Hubbard model only by effect of
irrelevant operators, it is likely that the ferromagnetic
instability sets in at a higher energy scale than that needed
for the charge instability to open up. 

The above considerations may depend, though, on the particular 
details of the microscopic model. For an extended Hubbard model
with nearest-neighbor interaction $V$, for instance, the 
assignment of bare couplings made from the nominal interactions 
in the model gives: $F_{intra \perp} \sim U + 4V$, 
$F_{inter \perp} \sim U + 4V$, $\tilde{F}_{inter \parallel}
\sim 8V $. In this case, the coupling driving the charge
instability becomes $F_{inter \perp} + 
\tilde{F}_{inter \parallel} - F_{intra \perp} -
\tilde{F}_{intra \parallel} \sim 8V$. It turns out that the model
with extended repulsive interaction $V$ falls in the 
universality class with the charge instability. The splitting
of the Van Hove singularity in the model has to be more
pronounced as the strength of the repulsive interaction $V$
is increased.

On the other hand, for $c' > c$, the tendency towards a
spin-density-wave instability prevails over any other
instability in the density of charge or spin. 
This is so because, in this regime, backscattering and Umklapp 
interactions are stronger than those considered in this 
section. Besides, the scaling equation governing the 
spin-density-wave instability has the same structure that 
Eqs. (\ref{z}) and (\ref{x}), 
but with the coefficient $c'$ instead of 
$c$ \cite{jpn}. This is also the region of the phase diagram 
in which $d$-wave superconductivity is likely to occur, due to 
the Kohn-Luttinger mechanism\cite{pin,kohn}. For
values of $t'$ below $\approx 0.276 t$, it is clear that the
energy scale of the formation of the Cooper pairs or the
spin-density-wave has to be much higher than the scale at which
the splitting of the Van Hove singularity may be a sensible
effect.

\section{Charge dynamics in contact with charge reservoir}

We have seen that a universal feature of our wilsonian RG scheme
is that the coupling to the total charge of the system is
renormalized to zero when the Fermi level approaches the Van
Hove singularity. This is just a consequence of the strong
screening processes that arise due to the divergent density of
states. That property does not have any sensible effect for a
closed system with constant number of particles, since the Fermi
energy can only be a monotonous function of the total charge.
However, if the system is instead at fixed chemical potential,
important effects may be derived from the mentioned result. The
description at fixed chemical potential has to do with the
situation in which the system is in contact with a charge
reservoir, that has a much larger content of particles and is
less susceptible of changes in its Fermi energy. This may be
also the relevant situation for the physics of the high-$T_c$
materials, regarding the interaction of the CuO layers with the
rest of the perovskite structure.

When the system does not have a fixed number of particles, the
Fermi energy displays a nontrivial dynamics which tends to pin
it to the Van Hove singularity\cite{mark,pin}. 
In the RG framework it has been
shown that the Fermi energy, taken as a running parameter
dependent on the high-energy cutoff, has a stable fixed-point
very close in energy to the Van Hove singularity\cite{pin}.
From the physical point of view, this leads to the important
consequence that a certain range of fillings should be forbidden
above and below the singularity. The prediction is that, for
nominal values of the chemical potential in that range, the
Fermi energy is led to the fixed-point at the singularity.
Only for lower or higher values of the chemical potential 
away from the region of attraction one may recover the regular 
evolution of the Fermi energy upon filling.

The pinning mechanism can be best understood by solving the
Schwinger-Dyson equation for the model in contact with a system
with large but constant density of states. The latter has then a
Fermi energy much less sensitive to changes in the total number
of particles, what amounts in practice to imposing the condition
of fixed chemical potential in the part of the system with the
Van Hove singularity. As in the previous section, the
Schwinger-Dyson equation written for the Fermi energy expresses
how the filling level for each of the systems is renormalized by
the shift of the one-particle levels to higher energies due to
the repulsive interaction. The self-consistent dependence of
this effect on the charge and the strength of the renormalized
interactions leads to the unconventional dynamics of the Fermi
level near the Van Hove singularity.

We model the system with the Van Hove singularity by taking the
density of states
\begin{equation}
n^{(1)} (\varepsilon) = - \frac{1}{\Lambda } \log (|\varepsilon
| /\Lambda ) \;  ,  -\Lambda < \varepsilon < \Lambda
\label{dos1}
\end{equation}
For the system with large but constant density of states, we
take a dependence of the form
\begin{equation}
n^{(2)} (\varepsilon) =  \frac{\alpha}{\Lambda } \;  ,  - \beta
\Lambda < \varepsilon
\label{dos2}
\end{equation}

We assume that, in the first system, the coupling to the total
charge, $\tilde{F}_{intra \parallel} + F_{intra \perp} +
\tilde{F}_{inter \parallel} + F_{inter \perp}$, 
is renormalized near the singularity according to Eqs.
(\ref{flow31}) and (\ref{flow32}). On the other hand, the
interaction between particles in the charge reservoir is
scale-independent and we suppose that it can be parametrized
by a constant coupling $F_0$ in the forward scattering channel.

As in the previous section, we introduce a common chemical
potential $\mu$ for the two systems, which enforces the
condition of thermodynamic equilibrium between them.  The
Schwinger-Dyson equation gives rise to the following pair of
nonlinear equations for the respective filling levels
$\varepsilon_{F1}$ and $\varepsilon_{F2}$ of the two systems
\begin{eqnarray}
\varepsilon_{F1} & = & \mu - \int^{\varepsilon_{F1}}_{-\Lambda} 
  d \varepsilon \; F (\varepsilon )  \; n^{(1)} (\varepsilon )
 - g_{\rm fwd} \int^{\varepsilon_{F2}}_{-\beta
\Lambda} d \varepsilon  \; n^{(2)} (\varepsilon )  \label{self21} \\
\varepsilon_{F2} & = & \mu - F_0 
  \int^{\varepsilon_{F2}}_{-\beta \Lambda} d \varepsilon \; n^{(2)}
(\varepsilon ) - g_{\rm fwd} \int^{\varepsilon_{F1}}_{-\Lambda}
d \varepsilon  \;  n^{(1)} (\varepsilon )
\label{self22}
\end{eqnarray}
where $F \equiv \tilde{F}_{intra \parallel} + F_{intra \perp} +
\tilde{F}_{inter \parallel} + F_{inter \perp}$ and
we have introduced a coupling constant $g_{\rm fwd}$ that
parametrizes the repulsion exerted on one of the systems by the
charge present in the other.

We stress once more that, in the above equations,
$\varepsilon_{F1}$ and $\varepsilon_{F2}$ are measured in the
reference frames in which the dependences $n^{(1)} (\varepsilon)
$ and $n^{(2)} (\varepsilon)$ are fixed by Eqs. (\ref{dos1}) and
(\ref{dos2}). As remarked in the previous section, the physical
picture is however that the one-particle levels are renormalized
to higher energy by a different amount in each of the systems,
so that both Fermi levels match at the end the common chemical
potential.

\begin{figure}
\begin{center}
\mbox{\epsfxsize 6.5cm \epsfbox{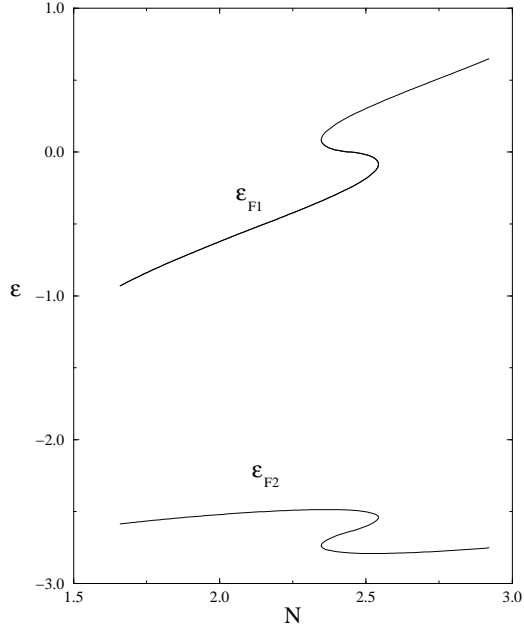}}
\end{center}
\caption{Self-consistent solutions for the respective filling
levels $\varepsilon_{F1}$ and $\varepsilon_{F2}$ in the system
with the Van Hove singularity and in the charge reservoir.}   
\label{twelve}
\end{figure}

The coupled set of equations (\ref{self21}) and (\ref{self22})
gives rise to nontrivial physical effects, as a consequence of
the nonlinearities introduced by the divergent density of states
$n^{(1)} (\varepsilon) $ and the renormalization of $F
(\varepsilon )$ close to the Van Hove singularity.  It is
interesting, for instance, to solve for the location of
$\varepsilon_{F1} $ and $\varepsilon_{F2} $ in terms of the
total charge $N$ in the two systems, given by
\begin{equation}
N = \int^{\varepsilon_{F1}}_{-\Lambda} d \varepsilon n^{(1)}
(\varepsilon ) + \int^{\varepsilon_{F2}}_{-\beta \Lambda} d
\varepsilon n^{(2)} (\varepsilon )
\end{equation}
The most remarkable effect is that there is not a one-to-one
correspondence between $N$ and the respective filling levels
$\varepsilon_{F1}$ and $\varepsilon_{F2}$.  The different
branches of the solution are represented in Fig. \ref{twelve}
for the particular values $F (\Lambda ) = F_0 = 4 \Lambda $,
$c/\pi^2 = 0.2$ and $g_{\rm fwd} = 3\Lambda $. The parameter $\beta $
has been chosen equal to $3.0$, and $\alpha $ has been set equal
to $4.0$, according to the idea of having a large density of
states in the second of the systems.

\begin{figure}
\begin{center}
\mbox{\epsfxsize 6.5cm \epsfbox{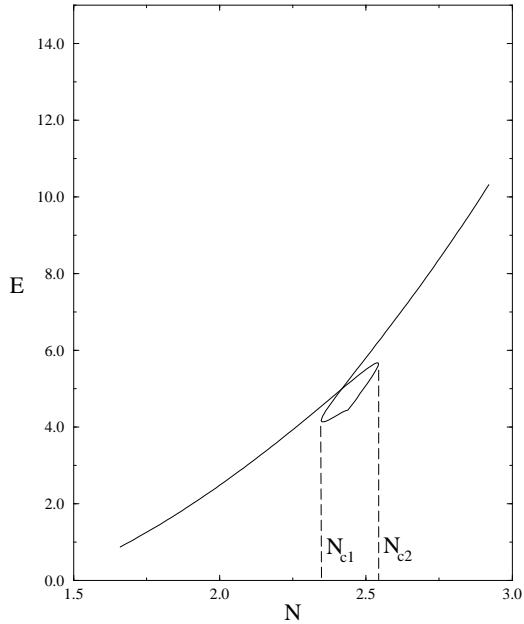}}
\end{center}
\caption{Total energy of the solutions shown in Fig.
\ref{twelve} versus total charge $N$.}
\label{thirteen}
\end{figure}

At low values of $N$, the filling of the first system with the
Van Hove singularity proceeds in a regular way, with a
monotonous increase of $\varepsilon_{F1} $. There is a point,
however, above which two other locations of $\varepsilon_{F1}$
become possible, closer to the singularity in the density of
states at $\varepsilon = 0$. 
In these instances, the corresponding filling level
$\varepsilon_{F2} $ in the second system suffers a 
decrease with respect to the expected value. It is interesting
to discern what of the possible solutions is most favorable
energetically. We have plotted in Fig. \ref{thirteen} the values
of the total energy $E$ versus the total charge $N$. We see that
the filling level closer to the Van Hove singularity
gives always the lowest-energy configuration of the system.

The result that turns out to be valid under very general
conditions is the existence of a certain range of filling levels
that are forbidden above and below the Van Hove singularity.  
This is in
agreement with previous analyses of the pinning of the Fermi
level of electrons near a Van Hove singularity\cite{mark,pin}.
The present study of the Schwinger-Dyson equation helps to
clarify the mechanism involved in that effect. It happens
that, for certain values of the charge $N$, this finds more
favorable to fill the Fermi sea up to the Van Hove singularity,
at the expense of the charge in the other system. In general,
there is a critical value $N_{c1} (N_{c2})$ of the total charge 
in which the filling level $\varepsilon_{F1}$ jumps discontinuously 
from the regular evolution upon adding (removing) particles  
to a position much
closer to the Van Hove singularity. This is in correspondence
with the onset of attraction to the stable fixed-point found
in the RG framework.

A last remark regarding the plot in Fig. \ref{thirteen} is that
the abrupt change in the lowest energy of the system at $N_{c1} 
(N_{c2})$ leads to phase separation for values of $N$ below 
(above) that critical value. It is clear, for instance, 
that for a certain range above $N_{c2}$  the whole system 
lowers its energy by splitting in two phases, one with a
higher value of the charge density and the other with the density
corresponding approximately to $N_{c2}$\cite{note2}. 
This reflects in another 
fashion that special stability is conferred to the system when
the Fermi level is at the Van Hove singularity.

\section{Conclusions}

In this paper we have adopted a wilsonian RG approach to
discern the charge instabilities of 2D electron systems at
the Van Hove filling. This kind of differential method of
renormalization was implemented in Ref. \onlinecite{rg1} to
discuss Fermi liquid theory in the context of the universality
classes of interacting fermion systems. When applied to the
system of electrons near the Van Hove singularity, we have seen
that the method leads to a rigorous analysis of the
instabilities in the spin and charge sectors.

It is well-known that the main problem
of dealing with the singular
density of states in the RG framework is that it gives rise to
harmful $\log^2 \Lambda $ divergences in the particle-particle
diagrams. These divergences cannot be removed by RG methods in a
standard fashion, as they actually point at the appearance of
nonlocal operators that are infrared divergent in the limit of
vanishing momentum at the singularity. In the differential RG
approach, however, a careful analysis of the kinematics shows
that, at least at the one-loop level, the divergences of the
particle-particle diagrams only affect the BCS channel\cite{rg1}.
The forward scattering
channel is only affected by conventional $\log \Lambda $
divergences. These are at the origin of potential instabilities
in the correlators for the density of spin and charge, that can
be now properly understood in the differential RG scheme.

An important property of our RG approach is that it allows to
preserve the spin rotational invariance at all the steps of the 
RG process. In fact, of all the flows that we have described in 
the space of couplings, there is only a very reduced number of 
combinations that realize the mentioned invariance.
This shows how stringent the wilsonian approach can be by 
enforcing symmetry constraints to determine the low-energy 
effective theory. 

Starting with bare repulsive interactions in the
forward and exchange channels, we have seen that there are only
two asymptotic low-energy behaviors consistent with the SU(2)
spin invariance.  One of them corresponds to the line
$\tilde{F}_{\parallel} = F_{\perp}$, for all the forward
scattering couplings, and $E_{\perp} = 0$, for all the exchange
couplings. Under these conditions, all the $F$ couplings are
renormalized to zero at low energies,
and it is clear from Eqs. (\ref{z}) and
(\ref{x}) that all the spin projections have equal dynamical
correlations at all points of the flow.  The other possibility
corresponds to the choice $F_{\perp} - \tilde{F}_{\parallel} =
E_{\perp}$ for all the couplings. In this case, these flow to a
strong coupling regime with singular response functions in the
spin sector. As discussed above, the $t-t'$ Hubbard model has a
low-energy behavior that falls within the latter class.

Turning to the charge instabilities, we have seen that the
interaction between the electrons at the two inequivalent saddle
points of the square lattice leads to two different universality
classes for 2D electron systems near a Van Hove singularity. 
One of them
corresponds to the RG flows below the bisector of the first
quadrant in Fig. \ref{ten}, for which $\tilde{F}_{ intra
\parallel} + F_{ intra \perp} > \tilde{F}_{ inter \parallel} +
F_{ inter \perp}$.  In this class, both combinations of
couplings are renormalized to zero at low energies, no response
function in the charge sector displays singular behavior, and
the instabilities may arise in the spin sector.

The other universality class corresponds to the
unstable flows with $\tilde{F}_{ intra \parallel} + F_{ intra
\perp} < \tilde{F}_{ inter \parallel} + F_{ inter \perp}$, which
lead to a singular response in the charge sector. We have shown
that this phase is characterized on physical grounds by the
splitting of the levels of the two inequivalent saddle points.
This kind of instability has been found recently in a numerical
RG study of the $t-t'$ Hubbard model\cite{hm}. 
A naive assignment of the bare couplings of the model gives 
$F_{ intra \perp} = F_{inter \perp} = U$ and 
$\tilde{F}_{ intra \parallel} = \tilde{F}_{inter \parallel} = 0$, 
placing it right at the boundary between the
two universality classes. However, the boundary is not itself
stable and the effect of any irrelevant perturbation may break
the balance in favor of either side. The findings of Ref.
\onlinecite{hm} show that this is indeed the case and that the
Hubbard model has to belong to the universality class with the
charge instability.  In any event, our analysis makes clear that
the singular response in the charge sector may develop before
any instability in the spin sector only for values of $t'$
above $\approx 0.276 t$. In that range, there is a competition
with the ferromagnetic instability that is also known to open up
in the model at the Van Hove filling\cite{sor}.

A feature common to both universality classes is that the
coupling to the total charge, $\tilde{F}_{ intra \parallel} +
F_{ intra \perp} + \tilde{F}_{ inter \parallel} + F_{ inter
\perp}$, vanishes in the low-energy limit. 
As a consequence of this fact we have seen that,
for an open system that is allowed to exchange particles with a
charge reservoir, there is a certain range of fillings forbidden
above and below the Van Hove singularity. This has the property
of attracting the Fermi level for the corresponding values of
the total charge in the mentioned range, as the system reaches
then its lowest energy when the Fermi energy is at the
singularity.

The mechanism of pinning to the Van Hove singularity could be
relevant to explain some of the properties of the hole-doped 
copper-oxide superconductors. Angle-resolved photoemission 
experiments\cite{photo}, as well as quantum Monte Carlo
computations for the $t-J$ and Hubbard models\cite{quantum}, 
have shown very flat portions of the quasiparticle dispersion at 
the boundary of the Brillouin Zone. In different compounds, the
Fermi level has been estimated to be very close to saddle
points of the band.  These observations have been contested by
the fact that the evidence for quasiparticles does not appear
quite clear, given the broad peak of the spectral weight near
the Fermi energy. However, the reduction in the quasiparticle
weight is another of the consequences which derives from the
interaction of electrons near a Van Hove singularity. It has
been shown that the electron wavefunction is strongly
renormalized in these circumstances\cite{np}.  
Although the quasiparticle description does not lose its 
validity, there is a strong attenuation of the quasiparticle 
pole as the Van Hove singularity is approached, and the 
normal state of the system adheres to the so-called marginal 
Fermi liquid behavior\cite{mfl}.

The greater stability attained when the Fermi level approaches
the Van Hove singularity could have experimental signatures
in other systems that are essentially two-dimensional and may
exchange particles with the environment. Interfaces like 
Sn/Ge(111) have been much studied recently, as they show a 
remarkable phase transition with the formation of a surface
charge-density-wave in the low-temperature phase\cite{cdw}. 
Photoemission experiments have shown the appearance of a 
very flat conduction band\cite{goldoni}, 
which is found at an energy sensibly smaller than predicted
by conventional band calculations. An important property is that
the system remains metallic accross the transition, what makes
plausible the description by means of weak coupling RG methods.
It has been proposed actually that the main features of the 
interface, including the loss of spectral weight below the
transition and the formation of the charge-density-wave
structure, can be explained by the effect of pinning at a Van
Hove singularity that is present in the conduction band of the
2D system\cite{br}. 

We remark finally that the stability of the Van Hove filling may
result in the effect of phase separation over a wide range
of nominal filling levels above and below the Van Hove
singularity\cite{note2}. 
This effect has to be realized when the system is 
in contact with a sufficiently large reservoir, as we have shown 
in the paper. A most important question would be to
ascertain, from the experimental point of view, to what extent
the perovskite structure of the high-$T_c$ materials may lead
to the pinning mechanism we have proposed, and whether the phase 
separation associated with it may bear some relation to that
observed in the form of stripes in the underdoped cuprates.

\vspace{0.5cm}
Useful discussions with F. Guinea are gratefully acknowledged.
This work has been supported by the CICYT grant PB96/0875
and the CAM (Madrid) grant 07/0045/98.


\begin{references}
\bibitem{note}
We note that the use of different conventions makes the
coupling $U_{back}$ in Ref. \onlinecite{jpn} to correspond to
the coupling $E_{inter \perp}$ defined in the present paper.
\bibitem{note3}
This value is also obtained from Ref. \onlinecite{lh} when the
proper normalization of the particle-hole susceptibility at 
vanishing momentum is taken into account.
\bibitem{note2}
Related ideas regarding phase
separation have been discussed in Ref. \onlinecite{mark}.
\end{references}
\end{document}